\documentclass{iopjournal}
\usepackage[normalem]{ulem}

\usepackage[backend=biber,
  style=authoryear,
  maxcitenames=1,   % show only the first author in citations
  mincitenames=1,
  maxbibnames=10,
  uniquename=false,   % don't add initials for disambiguation
  uniquelist=false
]{biblatex}
\DefineBibliographyStrings{english}{
  andothers = {\emph{et al}}
}
\renewbibmacro*{in:}{%
  \ifentrytype{article}
    {}
    {\printtext{\bibstring{in}\intitlepunct}}%
}

\begin{document}

\articletype{} %	 e.g. Paper, Letter, Topical Review...

\title{Microdosimetry Aspects in Diffusing Alpha-emitters Radiation Therapy. Part I: Effect of Broad Nucleus Size Distributions}

\author{Yevgeniya Korotinsky$^1$, Lior Arazi$^{1,*}$}

\affil{$^1$Unit of Nuclear Engineering, Faculty of Engineering Sciences, Ben-Gurion University of the Negev, Be'er-Sheva, Israel}

\affil{$^*$Author to whom any correspondence should be addressed.}

\email{larazi@bgu.ac.il}


\vspace{5mm}
\noindent Submitted to Physics in Medicine \& Biology on October 29, 2025

\begin{abstract}
\vspace{0.2cm}
\textbf{Objective:} Diffusing alpha-emitters Radiation Therapy (``Alpha DaRT'') is a new treatment modality focusing on the use of alpha particles against solid tumors. The introduction of Alpha DaRT in clinical settings calls for the development of detailed tumor dosimetry, which addresses biological responses such as cell survival and tumor control probabilities at the microscopic scale.\\ 
\vspace{0.1cm}
\textbf{Approach:} We present a microdosimetric model that links the macroscopic alpha dose, cell survival, and tumor control probability while explicitly accounting for broad distributions of spherical nucleus radii. The model combines analytic expressions for nucleus-hit statistics by alpha particles with Monte Carlo–based specific-energy deposition to compute survival for cells whose nucleus radii are sampled from artificial and empirically derived distributions. \\
\vspace{0.1cm}
\textbf{Main results:} Introducing finite-width nucleus size distributions causes survival curves to depart from the exponential trend observed for uniform cell populations, especially at high therapeutic doses. We show that the width of the nucleus size distribution strongly influences the survival gap between radiosensitive and radioresistant populations, diminishing the influence of intrinsic radiosensitivity on cell survival. Tumor control probability is highly sensitive to the minimal nucleus size included in the size distribution, indicating that realistic lower thresholds are essential for credible predictions.\\
\vspace{0.1cm}
\textbf{Significance:} Our findings highlight the importance of careful characterization of clonogenic nucleus sizes, with particular attention to the smallest nuclei represented in the data. We provide an extensive discussion on the origins of very small nuclei, addressing both genuine biological phenomena and methodological factors arising from histological reconstruction. Without addressing these small-nucleus contributions, tumor control probability may be substantially underestimated. Incorporating realistic nucleus size variability into microdosimetric calculations is a key step toward more accurate tumor control predictions for Alpha DaRT and other alpha-based treatment modalities. Methodological and biological advances will enable more reliable treatment planning and potentially enhance the clinical utility of microdosimetric models.
\end{abstract}
% ===========================================================
\section{Introduction}
\vspace{0.5cm}
The use of alpha-emitting isotopes for the treatment of cancer presents a promising approach supported by a compelling radiobiological rationale \parencite{MIRD_Pamphlet_22, Hall2018radiobiology}. Alpha-particles have a high linear energy transfer (LET) of $\sim60-230$~keV/$\mu$m in tissue, which produces a dense track of ionization within the cell nucleus, resulting in more complex DNA damage patterns than sparsely ionizing forms of radiation such as X/$\gamma$-rays and electrons (and, to some extent, also protons). These complex DNA lesions lead to clonogenic cell death with a high probability; in most cases, not more than a few alpha particle hits are required to this effect \parencite{Goodhead1999}, while the same biological endpoint would require the passage of $\sim10^3-10^4$ electrons through the nucleus \parencite{Goodhead1999,PougetJP}. 
The biological effect of alpha particles is only weakly dependent on the cellular oxygen level, making them equally effective against hypoxic and aerobic cells, and is also largely independent of the dose rate and cell-cycle stage \parencite{Bedford,Hall1972, Hall2018radiobiology}. Finally, due to their short range in tissue ($\sim40-90$~$\mu$m), the use of alpha particles can guarantee sparing surrounding healthy tissue, provided that the alpha emitters are brought to the immediate vicinity of the cancer cells. This appealing set of properties has led to multiple preclinical and clinical studies within the general framework of Targeted Alpha Therapy (TAT) \parencite{McDevitt2018, TAT_WG2018, Westrom2018,Tafreshi2019}, with the approved use of $^{224}$RaCl$_2$ treatments for bone metastases in castration-resistant prostate cancer \parencite{Shirley2014}. Due to the short range of alpha particles, TAT is generally considered to be more appropriate for the treatment of single cells and micrometastatic disease, although recent works show potential efficacy against macroscopic metastases \parencite{Sathekge2019}.

Contrary to TAT, Diffusing alpha-emitters Radiation Therapy (``Alpha DaRT'') focuses, from the outset, on the use of alpha particles against solid tumors. Its basic principle and underlying physics have been described in detail in \cite{Arazi2007, Arazi2020, Heger2023a, Heger2023b, Epstein2023, Heger2024, Dumancic2025}, and \cite{Zhang2025}. A large volume of publications has covered preclinical studies on mice-borne tumors as a stand-alone treatment \parencite{Arazi2007, Cooks2008, Cooks2009a, Lazarov, Cooks2012}, in combination with chemotherapy \parencite{Cooks2009, Horev2012, Mirlot2013, Reitkopf2015}, in combination with immunotherapy \parencite{Keisari2014, Confino2015, Confino2016, Domankevich2019, Domankevich2020, Keisari2020, Keisari2021,DelMare2023}, and in large animals \parencite{Sadoughi2024, Shoshan2025}. Lastly, clinical results are reported on in \cite{Bellia2019, Popovtzer2020, D_Andrea2023, Popovtzer2024}, and \cite{Miller2024}.

The introduction of Alpha DaRT in clinical settings calls for the development of detailed tumor dosimetry. Theoretical studies on Alpha DaRT have so far concerned macroscopic dose modeling through the time-dependent solutions of the Diffusion-Leakage (DL) model \parencite{Arazi2020}. In this work, we extend this theoretical framework by considering a complementary approach based on microdosimetry, which addresses biological responses such as cell survival and tumor control probabilities at the microscopic scale. Microdosimetry, founded on an original approach introduced by Harald H. Rossi \parencite{Rossi1955, Rossi1959, Rossi1960, Rossi1968, Rossi1997, Kellerer2002} and rooted in the stochastic nature of radiation interactions, accounts for statistical fluctuations in energy deposition in small targets such as cell nuclei, and their influence on survival probability and the probability of treatment success. Incorporation of microdosimetric properties into radiobiological modeling can potentially improve the prediction of observed variation in biological responses and the formulation of future treatment plans, as well as help evaluate existing clinical outcomes. This is particularly important for alpha-particle-based treatments, where the mean number of hits to cell nuclei can be small and subject to significant fluctuations. 

It has long been acknowledged that both the nucleus size and the intrinsic radiosensitivity of cells, \emph{$z_{0}$}, fundamentally affect the biological response of cells to radiation \parencite{Charlton1991,SR1999,Lazarov,Minguez2020}. Most microdosimetric studies consider uniform nucleus size populations (i.e. cells with discrete values of their nucleus radii). In contrast, continuous distributions of cell and nucleus sizes have been largely ignored, in part due to lack of robust techniques and quantitative models describing the tumor cells/nucleus geometry. While studying uniformly-sized nuclei can be informative, it represents an idealized biological configuration. Realistically, cell and nucleus sizes can significantly vary within a given tumor, depending on the tumor and host tissue types, cellular microenvironment, and cell-cycle stage.

In this work, we explore the impact of continuous nucleus size distributions on cell survival and tumor control probability. We present a microdosimetric analysis that links macroscopic dose calculations and cell survival estimates for spherical cell nuclei of varying sizes (drawn from broad distributions) and intrinsic radiosensitivity. The effect is first explored by considering an artificial Gaussian distribution of nucleus radii, and subsequently using a broad distribution derived by Poole et al. from histological data \parencite{Poole2015}. This study represents the initial installment in a forthcoming series of microdosimetric analyzes related to Alpha DaRT, which will further discuss the effects of local dose gradients and implement complete tumor control probability (TCP) calculations. Although the present analysis is framed within the context of Alpha DaRT, its findings have broader implications for other alpha-particle-based treatment modalities. 
% ===========================================================

% ===========================================================
\section{Methods}
\vspace{0.5cm}
%--------------------------------------------------------------
\subsection{General framework}

For this study, we developed a microdosimetric model based on analytical and Monte Carlo (MC) calculations implemented in MATLAB. The model considers a tumor region subject to a macroscopic alpha-particle dose $D_{\alpha}$. The initial energies of the alpha particles are selected, with suitable branching ratios, from the line spectrum of the alpha-emitters released from the Alpha-DaRT source ($^{220}$Rn, $^{216}$Po, $^{212}$Bi, $^{212}$Po) \parencite{Nudat3}. The region's size is set to include all alpha particle tracks that can potentially hit the nucleus (we subsequently refer to this region as the ``volume of interest'', VOI).

\begin{figure}
   \begin{center}
   \includegraphics[width=\textwidth]{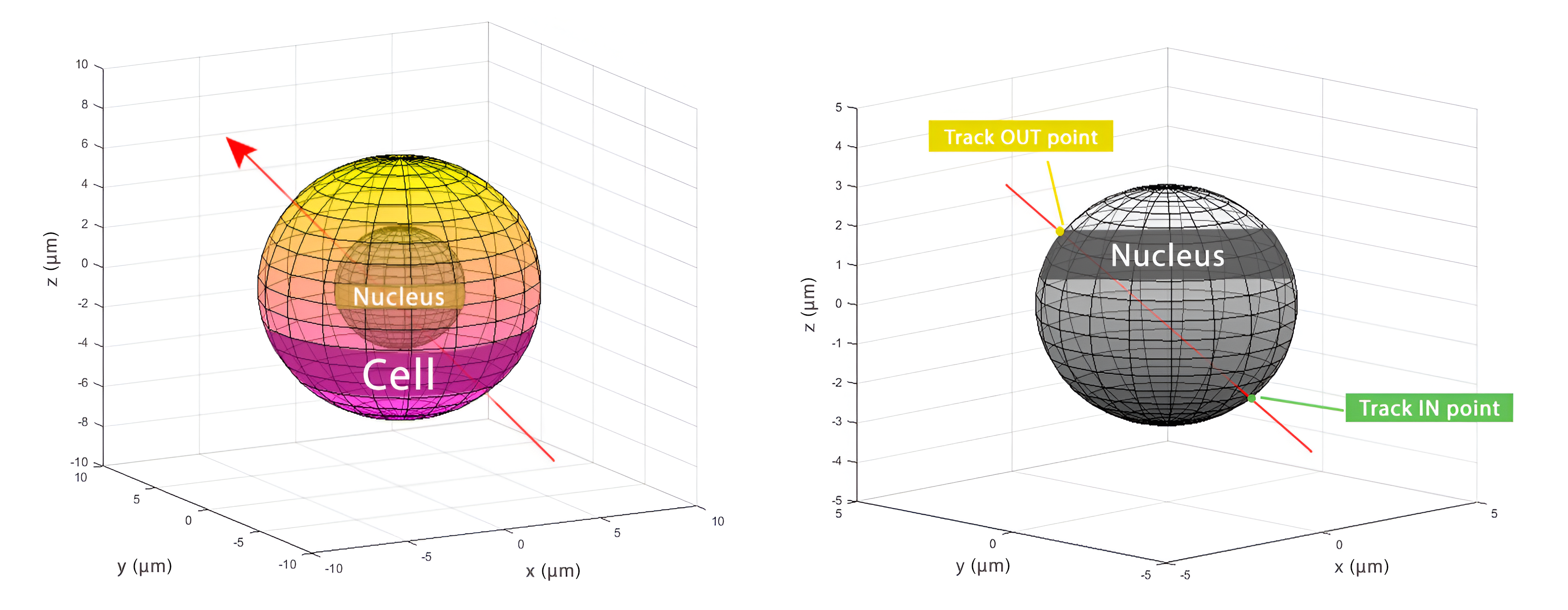}
   \caption
   {Visualization of the simulated target. (Left) A spherical cell containing a nucleus crossed by an alpha particle track marked by a red arrow. (Right) Zoom-in view on the nucleus with the hit points marked by captions.
   \label{fig:Cell_model} 
   }
    \end{center}
\end{figure}

Cells and nuclei are modeled as two concentric spheres of liquid water (Figure \ref{fig:Cell_model}). We focus on a single cell and assume that the macroscopic dose is uniform in the VOI (the effect of dose gradients will be considered in a separate publication). The VOI dose is proportional to the mean number of alpha decays in this volume, as given by:

\begin{equation}
	D_{\alpha} = N_{decays}\cdot\frac{E_{\alpha}}{\rho\cdot VOI} \label{eq:D_macro1}
\end{equation}
where $\rho$ is the tissue density (taken as 1.0 g/cm$^3$) and $E_{\alpha}$ is the energy of the alpha particles. The VOI was modeled as a concentric sphere of radius $R_n + R_{\alpha}$, where $R_n$ is the nucleus radius and  $R_{\alpha}$ is the continuous-slowing-down-approximation (CSDA) range of the alpha particle in water. 

Alpha particles are assumed to follow straight lines with no straggling. Tracks are classified as ``hits'' if they cross the nucleus envelope at least once (at a maximal distance $R_{\alpha}$ from their starting point). We used NIST ASTAR for the energy-dependent stopping power $dE/dX$ \parencite{ASTAR} to derive tabulated data representing the energy $E(r)$ of an alpha particle starting with an initial energy $E_0$ as a function of the radial distance from its emission point, gradually dropping to zero at the CSDA range corresponding to $E_0$. The model employed here identifies the intersection points of simulated tracks with a sphere mimicking the nucleus (see Figure \ref{fig:Cell_model}, right panel). Utilizing the coordinates of the track ``IN'' and ``OUT'' points, the particle's energy upon entry and exit from the nucleus is determined by interpolating the tabulated values of $E(r)$. If the track crosses the nucleus at two points, the deposited energy is $E_{dep}=E(IN)-E(OUT)$. If the track starts inside the nucleus $E_{dep}=E_0-E(OUT)$ and if the track stops inside the nucleus $E_{dep}=E(IN)$. 

For the simple case of a uniform spatial distribution of alpha emitters (inside the nucleus, cytoplasm, and extracellular space), the mean number of hits can be determined analytically\footnote{In fact, an analytic expression can be readily derived for any spherically-symmetric source distribution.} by dividing the VOI into concentric spherical shells of radius $r$ with infinitesimal width $dr$. Each shell encloses all possible alpha emission points within a distance $[r; r + dr]$ from the nucleus center. The probability of an alpha particle emitted from an isotropic point source at a distance $r$ from the nucleus center to hit the nucleus can be shown to satisfy:

\begin{equation}
P(r) =
\left\{
	\begin{array}{ll}
		1  & \mbox{} r \leq R_{n} \\\\
		\frac{1}{2}\left(1 - \sqrt{1 - \frac{R_{n}^2}{r^2}}\right) & \mbox{} R_{n} < r \leq \sqrt{R_{n}^2 + R_{\alpha}^2} \\\\
		\frac{1}{2}\left(1 - \frac{r^2 + R_{\alpha}^2 - R_{n}^2}{2R_{\alpha}r}\right) & \mbox{} \sqrt{R_{n}^2 + R_{\alpha}^2} < r \leq R_{n} + R_{\alpha} 
		
	\end{array}
\right.   \label{eq:hit_probability}
\end{equation}

\noindent The average number of total hits is then calculated by the integral

 \begin{equation}
	\overline{n}_{hit} = \int_{0}^{R_{n} + R_{\alpha}} n_{decays}P(r)4 \pi r^2  \,dr = D_{\alpha}\frac{\rho}{E_{\alpha}}\left(\frac{4\pi}{3}R_{n}^3 + R_{\alpha} \cdot \pi R_{n}^2\right)    \label{eq:n_hit}
\end{equation}
where $n_{decays}=N_{decays}/VOI$ is the number density of the decays, and the relation between $N_{decays}$ and $D_{\alpha}$ is given in Equation (\ref{eq:D_macro1}). The mean number of nucleus hits is proportional to the macroscopic dose and has two contributions \parencite{SR1995}: the first term comes from alpha decays occurring inside the nucleus and is therefore proportional to its volume. The second term is proportional to the nucleus cross-section and corresponds to alpha emissions from the outside. Since $R_{\alpha}$ is much larger than $R_n$, this is the dominant contributor to the number of hits.

\begin{figure}
 	\begin{center}
 		\includegraphics[scale=0.7]{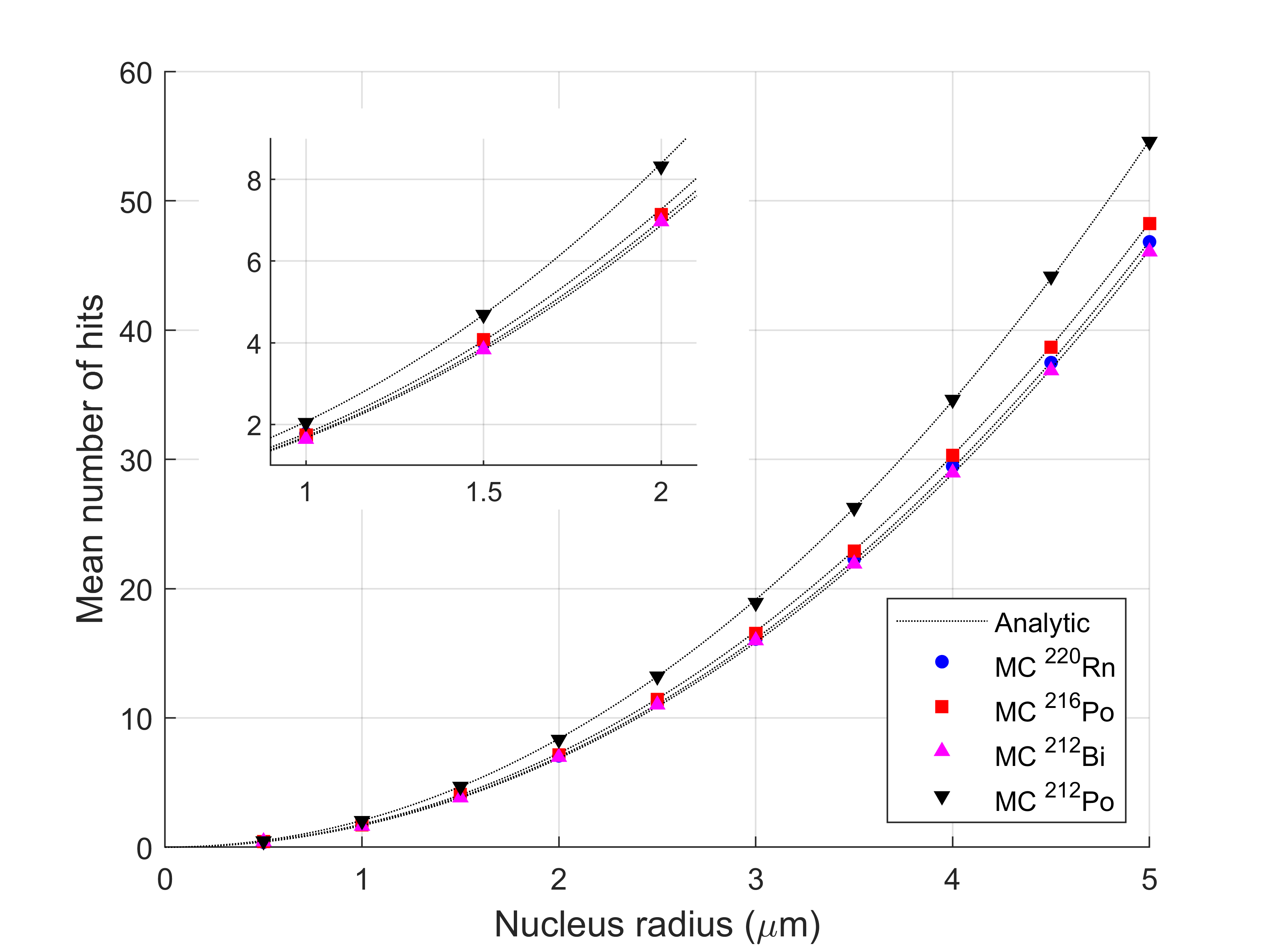}
 		\caption{Monte-Carlo calculation of the mean number of alpha-particle hits to the nucleus as a function of its radius, resulting from uniformly distributed alpha emitters: $^{220}$Rn (6.29 MeV), $^{216}$Po (6.78 MeV), $^{212}$Bi (6.05, 6.09 MeV) and $^{212}$Po (8.79 MeV), each for an absorbed dose of 10~Gy. Dotted lines represent the respective analytical calculation according to Equation (\ref{eq:n_hit}). The inset focuses on the region of small radii.} \label{fig:n_hit_Rn}
 	\end{center}
 \end{figure}

The actual number of hits to a given nucleus exposed to a macroscopic dose $D_{\alpha}$ is governed by Poisson statistics. Thus, in addition to the analytical calculation of the mean number of hits, we used the MC model to generate frequency distributions of the actual number of hits. The dependence of the mean number of hits to the nucleus on the nucleus radius---for all alpha emitters and an absorbed dose of 10~Gy---is displayed in Figure \ref{fig:n_hit_Rn}, which shows both the analytical expression (Equation \ref{eq:n_hit}) and the average values obtained from the MC simulation with 1000 repetitions for each choice of nucleus radius.  

\begin{figure}[htbp!]
 	\begin{center}
 		\includegraphics[width=\textwidth]{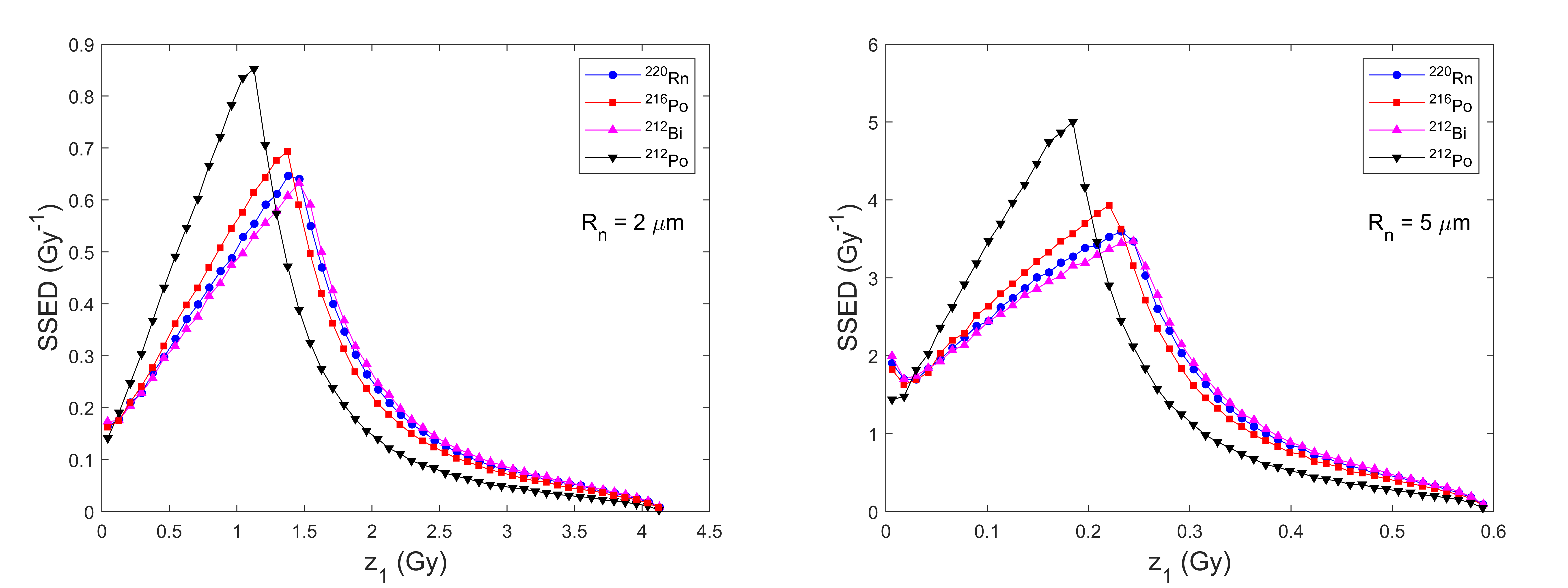}
 		\caption{Monte-Carlo calculation of the Single-hit Specific Energy Distribution (SSED) for $^{220}$Rn, $^{216}$Po, $^{212}$Bi, and $^{212}$Po, for a spherical nucleus with a radius of 2~$\mu$m (left) and 5~$\mu$m (right). The alpha-emitters are uniformly distributed inside and outside of the nucleus.} \label{fig:SSEDs}
 	\end{center}
 \end{figure}

In a microdosimetric description, the macroscopic dose is replaced by the \textit{specific energy} \emph{z}, defined as the ratio between the total deposited energy imparted by multiple hits $\epsilon$ to the target and the target mass \emph{m}: $z=\epsilon/m$. In our context, the specific energy per a single passage is $z_1 = E_{dep}/M_{nuc}$, where $M_{nuc}$ is the nucleus mass. For a given absorbed dose (by a specific alpha emitter) and a given nucleus size, the simulation is repeated numerous times to produce the statistical distribution of the Single-hit Specific Energy Depositions (SSED), $f_1(z_1;R_n,E_{\alpha})$. Figure \ref{fig:SSEDs} shows the MC-calculated isotope-specific SSEDs for a nucleus radius of 2~$\mu$m and 5~$\mu$m. Note that the single-hit specific energy for the larger nucleus is $\sim6$-fold smaller, as it roughly scales as $1/R_n^2$.

%--------------------------------------------------------------
\subsection{Nucleus size distributions}

\begin{figure}
   \begin{center}
   \includegraphics[width=0.8\textwidth]{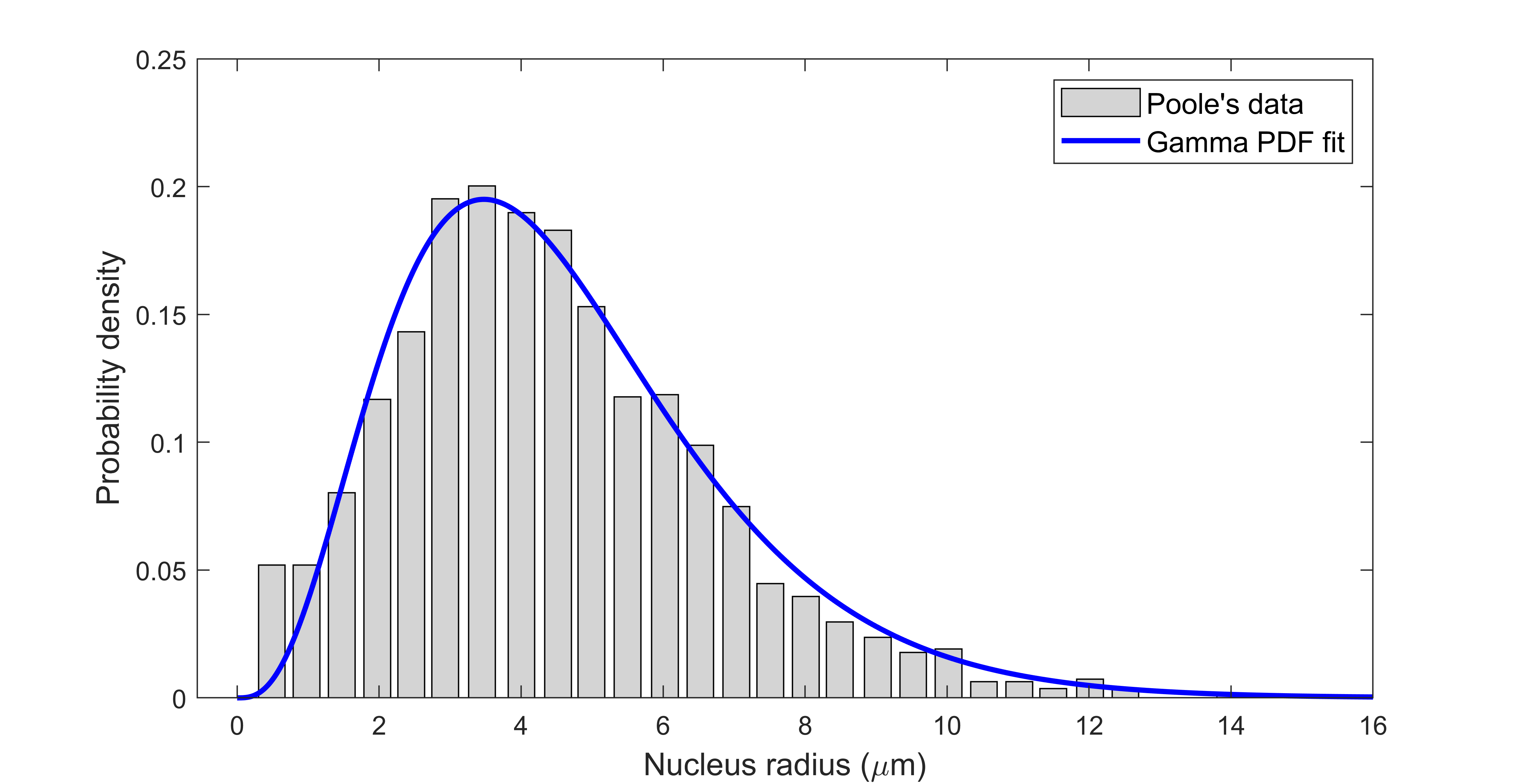}
   \caption{Nucleus radius distribution data extracted from Poole et al. 2015, fitted using a Gamma distribution function with parameters $k$ = 4.047 and $\theta$ = 1.141 with no lower threshold on the nucleus radius.
   \label{fig:Shirin_dist} 
   }
    \end{center}
\end{figure}

Cells inside a tumor can have varying nucleus sizes, which can have a strong effect on cell survival. Here, we investigate this effect starting with an artificial Gaussian distribution of nucleus radii and then using an empirical broad distribution. For the Gaussian probability distribution function (PDF) we adopt a mean radius of 4.44~$\mu$m (same as the average value of the empirical distribution discussed below) and a varying standard deviation $\sigma$. As for empirical distributions, experimental data on in-vivo nucleus size distributions are scarce to non-existent. The only dataset we are aware of is that reported in Poole, Ahnesjö, and Enger's study from 2015 \parencite{Poole2015} which introduced an algorithm able to reconstruct 3D cellular and nuclear geometries based on measured 2D histology samples (the data sample was provided for primary breast and prostate cancer cells). Poole's data are shown in Figure \ref{fig:Shirin_dist}. The raw data were fitted using a Gamma probability distribution function, which closely approximates the observed experimental profile. The Gamma distribution was parameterized by shape and scale parameters, denoted as $k$ and $\theta$, respectively:

\begin{equation}
    f(R_n)=\frac{1}{\Gamma(k) \theta^k}(R_n-R_{th})^{k-1}\;e^{-(R_n-R_{th})/\theta} \label{eq:Gamma_pdf_with_threshold}
\end{equation}
where $\Gamma(k)$ is the Gamma function and $R_{th}$ is a threshold nucleus radius, below which the probability distribution function is set to zero. We introduced this threshold because Poole's data extend to vanishingly small radii, which is a feature to be discussed in greater detail in the following. For both distributions (Gaussian and empirical), a total of $10^7$ random values of $R_n$ were obtained by inverse transform sampling.

%--------------------------------------------------------------
\subsection{Survival probability modeling}

For the case of a uniform spatial distribution of alpha emitters (inside the nucleus, cytoplasm, and extracellular space), the mean number of hits can be determined analytically using Equation \ref{eq:n_hit}. The overall survival probability within a population of identical cells subjected to a macroscopic absorbed alpha dose $D_{\alpha}$ is given by \parencite{Rossi1968, Kellerer1970, Roesch1977, SR1992}:
\begin{equation}
	SP_{micro}(D_{\alpha};R_n,E_{\alpha},z_0) = \exp\left[-\overline{n}_{hit}(D_{\alpha};R_n,E_{\alpha})\cdot(1 - T_{1}(z_{0};R_n,E_{\alpha}))\right]   \label{eq:mic_SSED_cell_survival}
\end{equation}
where $z_0$ is the intrinsic radiosensitivity of the cell (corresponding to the specific energy leading to a single cell survival probability of 37\%), and $T_1(z_0;R_n,E_{\alpha})$ is the Laplace transform of the SSED, $f_1(z_1;R_n,E_{\alpha})$:\\
\begin{equation}
	T_1(z_0;R_n,E_{\alpha}) = \int_{0}^{\infty} f_1(z_1;R_n,E_{\alpha})\exp(-z_1/z_0) ~dz_1    \label{eq:T1}
\end{equation}

For a population of cells, whose nucleus size is sampled from a PDF, the survival probability is calculated individually for each cell, using Equation \ref{eq:mic_SSED_cell_survival}, and the overall surviving fraction is calculated as the sum of all individual surviving probabilities divided by the number of cells.

\begin{figure}
 	\begin{center}
 		\includegraphics[width=\textwidth]{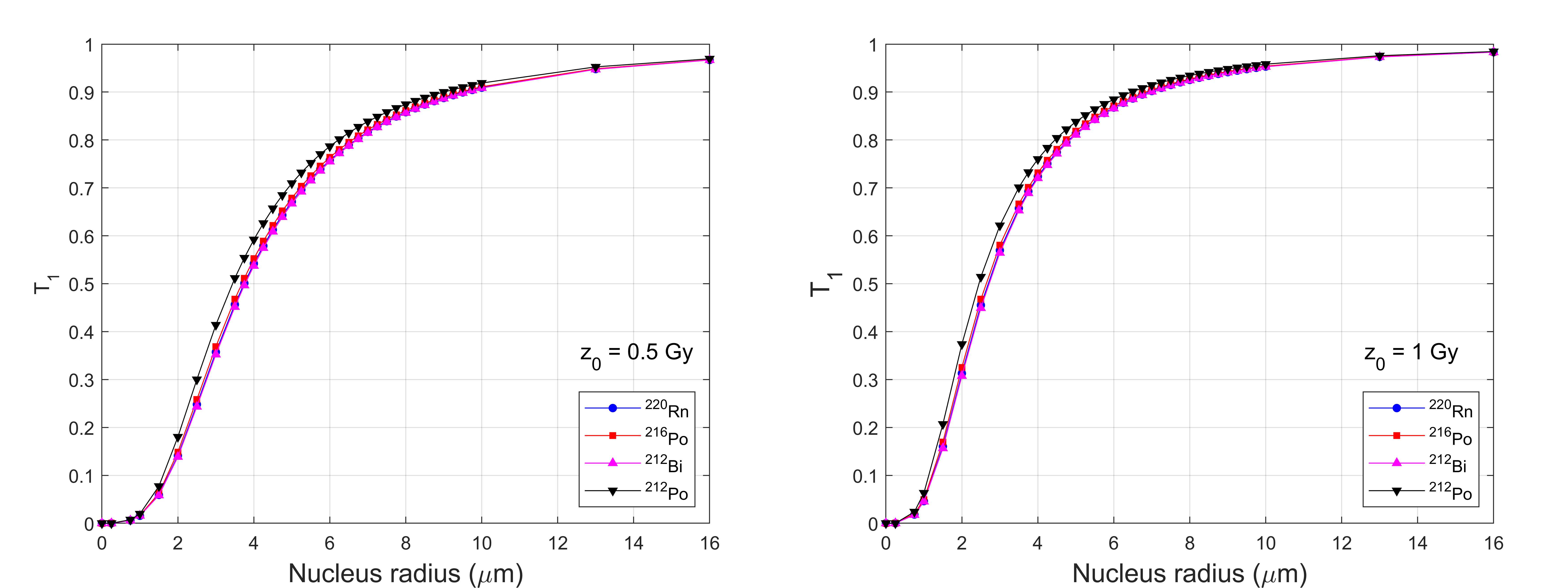}
 		\caption{Laplace transforms $T_1(z_0;R_n,E_{\alpha})$ of the SSED for $^{220}$Rn, $^{216}$Po, $^{212}$Bi, and $^{212}$Po, for a range of nucleus radii, and two selected values of the radiosensitivity parameter: $z_0=0.5$~Gy (left) and $z_0=1$~Gy (right). The alpha-emitters are uniformly distributed inside and outside of the nucleus.} \label{fig:T1s}
 	\end{center}
\end{figure}

For computational efficiency, we used our MC simulation to create a dataset of SSEDs and tabulated values for $T_1(z_0;R_n,E_{\alpha})$ for all relevant alpha-particle energies, for a wide range of values for the nucleus radii, and for selected values of $z_0$. When calculating the survival probability of a cell with a particular nucleus radius $R_n$, the corresponding value of $T_1(z_0;R_n,E_{\alpha})$ was found by interpolation on the pre-calculated dataset. Figure \ref{fig:T1s} shows $T_1$ as a function of nucleus radius for all the diffusing alpha-emitter daughters for $z_0=0.5$~Gy and $z_0=1$~Gy. The figure shows a weak dependence of $T_1(R_n)$ on the alpha-particle energy, with a modest difference between the two selected values for the radiosensitivity parameter.

%--------------------------------------------------------------
\subsection{Tumor control probability modeling}

In macroscopic tumors, only a tiny fraction of cells can proliferate indefinitely and produce large colonies of secondary cells \parencite{Pollak1984,Chapman2014}. Such cells are called ``clonogens'', and the ultimate target of radiation therapy is their complete eradication. The associated quantity is the \textit{Tumor Control Probability} (TCP), defined as the probability that treatment of a given tumor results in zero surviving clonogens. Since cell killing is a stochastic process, one can imagine multiple repetitions of a treatment consisting of some radiation dose distribution to a given clonogen population. We denote by $\overline{N}$ the average number of surviving clonogens over many such repetitions. Assuming that the number of surviving clonogens follows a Poisson distribution and that complete eradication of the tumor requires zero surviving clonogens, the tumor control probability is:
\begin{equation}
	TCP = P(0;\overline{N})=\exp(-\overline{N})     \label{eq:TCP}
\end{equation}

To illustrate the effect of sampling nuclei sizes from a broad distribution on the TCP, we limit the present analysis to the case of a uniform absorbed dose, deferring the discussion of realistic Alpha-DaRT source lattices to a separate publication. A total of $N_0$ clonogens are irradiated by the four daughter alpha emitters in the decay chain of $^{224}$Ra. We start with the total alpha dose $D_{\alpha,tot}$ and the $^{212}$Pb leakage probability $P_{leak}(\mathrm{Pb})$ (i.e., the probability that a $^{212}$Pb atom released or created inside the tumor leaks out of it through the blood before it decays \parencite{Arazi2007}). From these two values, we find for each isotope its individual alpha dose, from which---using Equation \ref{eq:n_hit}---we calculate the associated mean number of nucleus hits for a given nucleus radius.

The total alpha particle dose is made up of contributions from pair isotopes,  $^{220}\mathrm{Rn}+^{216}\mathrm{Po}$ and $^{212}$Bi/$^{212}$Po:
\begin{equation}
    D_{\alpha,tot}=D_{\alpha}(\mathrm{RnPo})+D_{\alpha}(\mathrm{BiPo}),
\end{equation}
where:
\begin{equation}
    D_{\alpha}(\mathrm{RnPo})=\frac{n_{decays}(\mathrm{Rn})}{\rho}\left(E_{\alpha}(\mathrm{Rn})+E_{\alpha}(^{216}\mathrm{Po})\right)   
\end{equation}

\begin{equation}
    D_{\alpha}(\mathrm{BiPo})=\frac{n_{decays}(\mathrm{Bi})}{\rho}\left(0.36\overline{E}_{\alpha}(\mathrm{Bi})+0.64E_{\alpha}(^{212}\mathrm{Po})\right)  
\end{equation}

\begin{equation}
    n_{decays}(\mathrm{Bi})=(1-P_{leak}(\mathrm{Pb}))\,n_{decays}(\mathrm{Rn})
\end{equation}
where $n_{decays}(\mathrm{Rn})$ and $n_{decays}(\mathrm{Bi})$ are the total number of decays of $^{220}$Rn and $^{212}$Bi per unit volume. The ratio between the two dose components $\eta=D_{\alpha}(\mathrm{BiPo})/D_{\alpha}(\mathrm{RnPo})$ is therefore:

\begin{equation}
    \eta = \frac{\left(1-P_{leak}(\mathrm{Pb})\right)\cdot\left( 0.36\overline{E}_{\alpha}(\mathrm{Bi})+0.64E_{\alpha}(^{212}\mathrm{Po})\right)}{E_{\alpha}(\mathrm{Rn})+E_{\alpha}(^{216}\mathrm{Po})}
\end{equation}

\noindent And the isotope-specific alpha doses are:

\begin{equation}
    D_{\mathrm{Rn}} = \frac{D_{\alpha,tot}}{(1+\eta)\left(1 + E_{\alpha}(\mathrm{^{216}Po})/E_{\alpha}(\mathrm{Rn})\right)} 
   \label{eq:Rn_breakdown}
\end{equation}

\begin{equation} 
    D_{\mathrm{Po216}} = \frac{E_{\alpha}(\mathrm{^{216}Po})}{E_{\alpha}(\mathrm{Rn})} \cdot D_{\mathrm{Rn}}
   \label{eq:Po216_breakdown}
\end{equation}

\begin{equation}
    D_{\mathrm{Bi}} = \frac{\eta\cdot D_{\alpha,tot}}{(1+\eta)\left(1 + 64/36\cdot E_{\alpha}(\mathrm{^{212}Po})/\overline{E}_{\alpha}(\mathrm{Bi})\right)}
   \label{eq:Bi_breakdown}
\end{equation}

\begin{equation}
    D_{\mathrm{Po212}} = \frac{64}{36}\cdot \frac{E_{\alpha}(\mathrm{^{212}Po})}{\overline{E}_{\alpha}(\mathrm{Bi})}\cdot D_{\mathrm{Bi}}
   \label{eq:Po212_breakdown}
\end{equation}

Each clonogen is assigned a nucleus radius $R_n$ sampled from a size distribution and a fixed radiosensitivity parameter $z_0$. For each clonogen, the survival probabilities corresponding to each isotope are determined separately from the respective alpha dose and mean number of alpha hits (Equation \ref{eq:n_hit}) and from the associated SSED, using the Laplace-transform expression (Equation \ref{eq:mic_SSED_cell_survival}). For simplicity, no low-LET contribution to cell killing is assumed in the present analysis. The clonogen compound survival probability is calculated as a product of the four survival factors, assuming independent DNA damage by all particles hitting the nucleus:
\begin{equation}
    SP_i = SP_{i,\alpha}(\mathrm{Rn})\cdot SP_{i,\alpha}(^{216}\mathrm{Po})\cdot SP_{i,\alpha}(\mathrm{Bi})\cdot SP_{i,\alpha}(^{212}\mathrm{Po})     \label{eq:SP_total}
\end{equation}
where $i$ is an index specifying the clonogen. The average number of surviving clonogens is estimated as:
\begin{equation}
    \overline{N}(D_{\alpha,tot};P_{leak}(\mathrm{Pb}))=\sum_{i=1}^{N_0} SP_i
\end{equation}
and the TCP as a function of the total alpha dose is found using Equation \ref{eq:TCP}.

%===============================================================
\section{Results}
\vspace{0.5cm}
%--------------------------------------------------------------
\subsection{Survival probability: artificial Gaussian distribution}

\begin{figure}[ht!]
   \begin{center}
   \includegraphics[width=\textwidth]{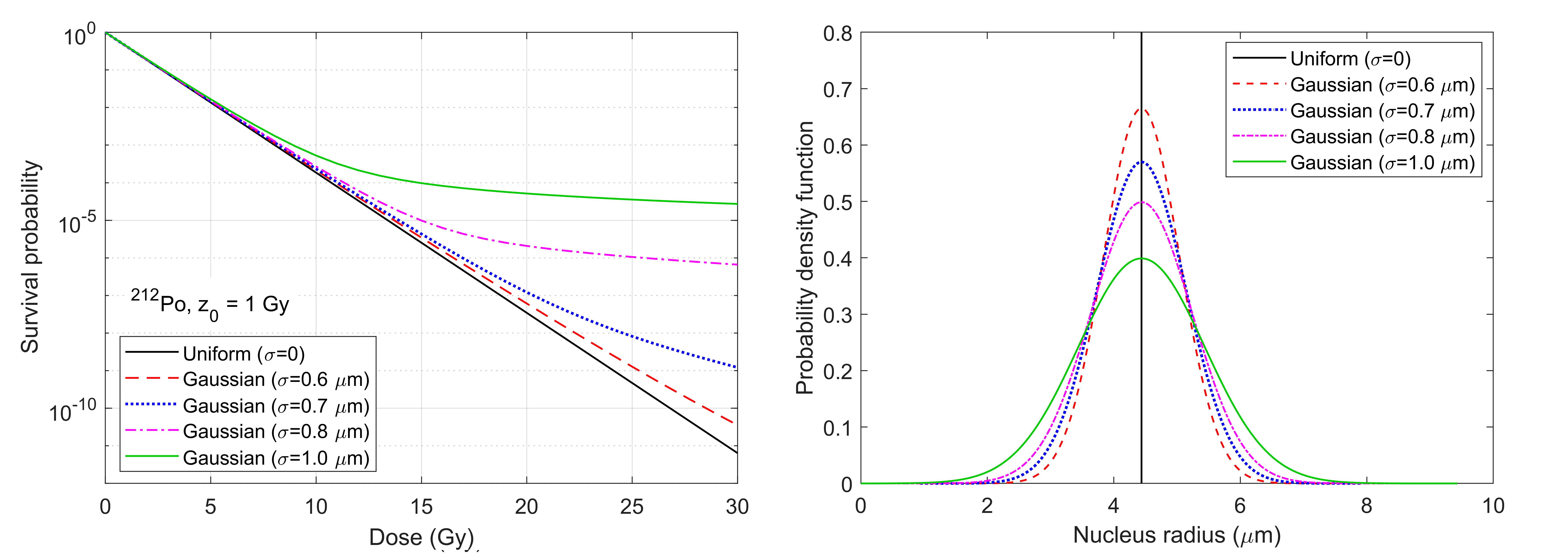}
   \caption{(Left) Survival probability of cells with nucleus radii sampled from two types of distributions: uniform and Gaussian with increasing standard deviation values. The cells are exposed to $^{212}$Po alpha particles and have a radiosensitivity $z_0=1$~Gy. All distributions maintain a mean nucleus radius of 4.44 $\mu$m. (Right) Associated nucleus radius distributions.
   \label{fig:SP_Gauss_width} 
   }
    \end{center}
\end{figure}

Figure \ref{fig:SP_Gauss_width} shows the survival curves of cells whose nucleus radii are sampled from Gaussian distributions of varying widths, as a function of the dose delivered by $^{212}$Po alpha particles, with a radiosensitivity parameter $z_0=1$~Gy. Incorporating a nucleus size distribution with a finite width leads to survival curves that depart from a strictly exponential trend. At lower doses typical of in-vitro studies (up to a few Gy), the curves closely resemble exponential behavior. 

\begin{figure}[ht!]
   \begin{center}
   \includegraphics[width=\textwidth]{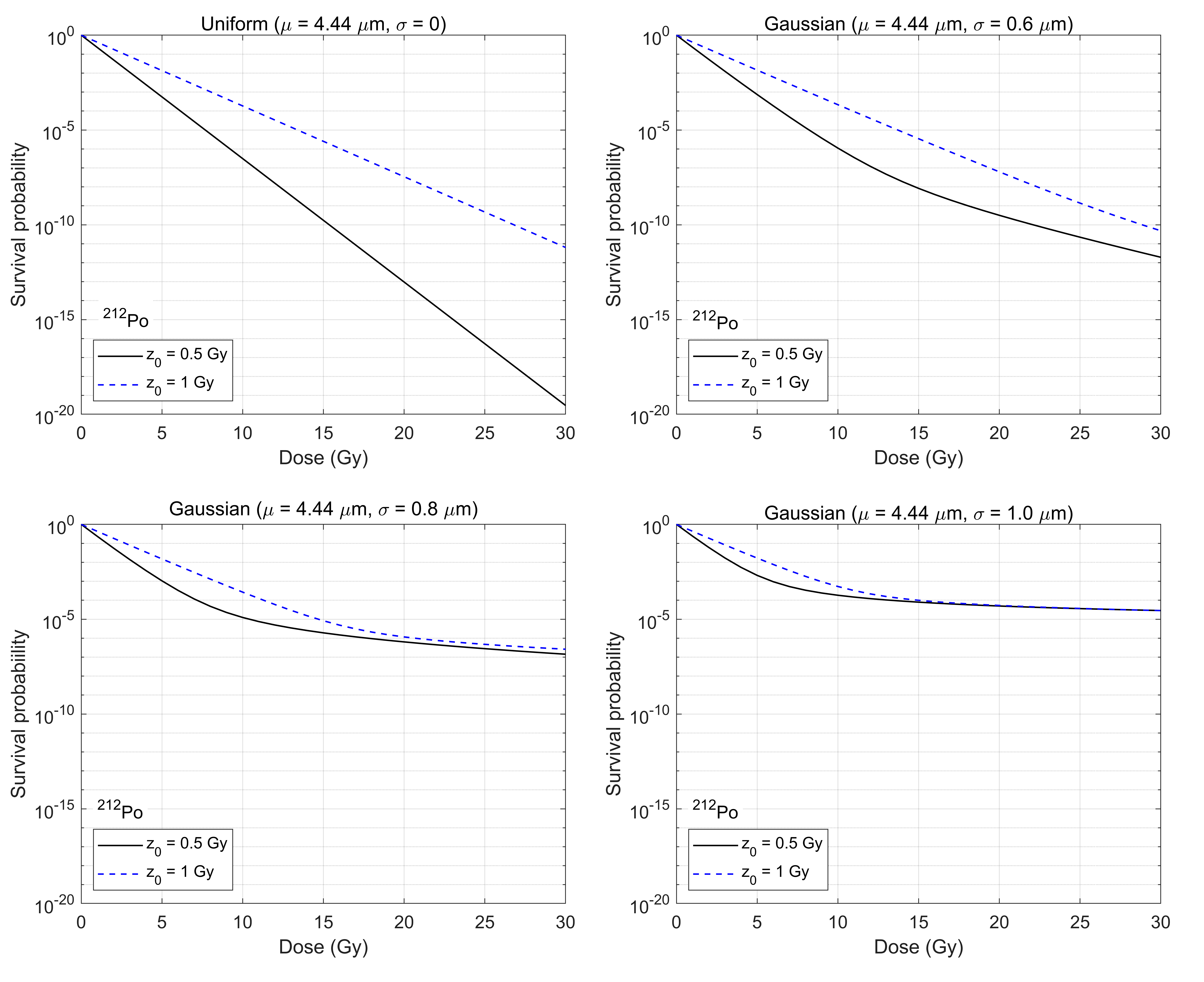}
   \caption{Survival probability plotted against dose for nucleus radii distributed uniformly and according to a Gaussian distribution with increasing width, both with a fixed mean of 4.44 $\mu$m. Each panel features two curves representing radiosensitive ($z_0=0.5$~Gy) and radioresistant ($z_0=1$~Gy) cells.
   \label{fig:SP_z0_diff} 
   }
    \end{center}
\end{figure}
However, at higher doses prevalent in tumor treatment, the deviation from the exponential trend intensifies noticeably. In distributions of nucleus radii with finite non-zero widths, an increase in dose results in larger cells succumbing first, leaving cells with smaller nucleus sizes alive. The diminishing nucleus radius corresponds to a gentler slope in the survival curves, reflecting the increasing influence of these smaller nuclei on the observed deviation.
\newpage
We further examined the impact of the width of the nucleus radius distribution on the survival disparity between cells with two radiosensitivity $z_0$ values: 0.5 and 1~Gy (reported values of $z_0$ typically lie in the range $\sim0.25-1.55$~Gy \parencite{Charlton1996, Lazarov}). The results are illustrated in Figure \ref{fig:SP_z0_diff}, showing that as the width of the nucleus size distribution increases, the gap between the survival curves corresponding to radiosensitive ($z_0$ = 0.5 Gy) and radioresistant cells ($z_0$ = 1 Gy) narrows down. In other words, as the fraction of cells with small nuclei included in the initial population increases, the influence of $z_0$ on survival decreases.
%--------------------------------------------------------------
\subsection{Survival and tumor control probabilities: broad nucleus size distribution}
 
\begin{figure}
   \begin{center}
   \includegraphics[width=\textwidth]{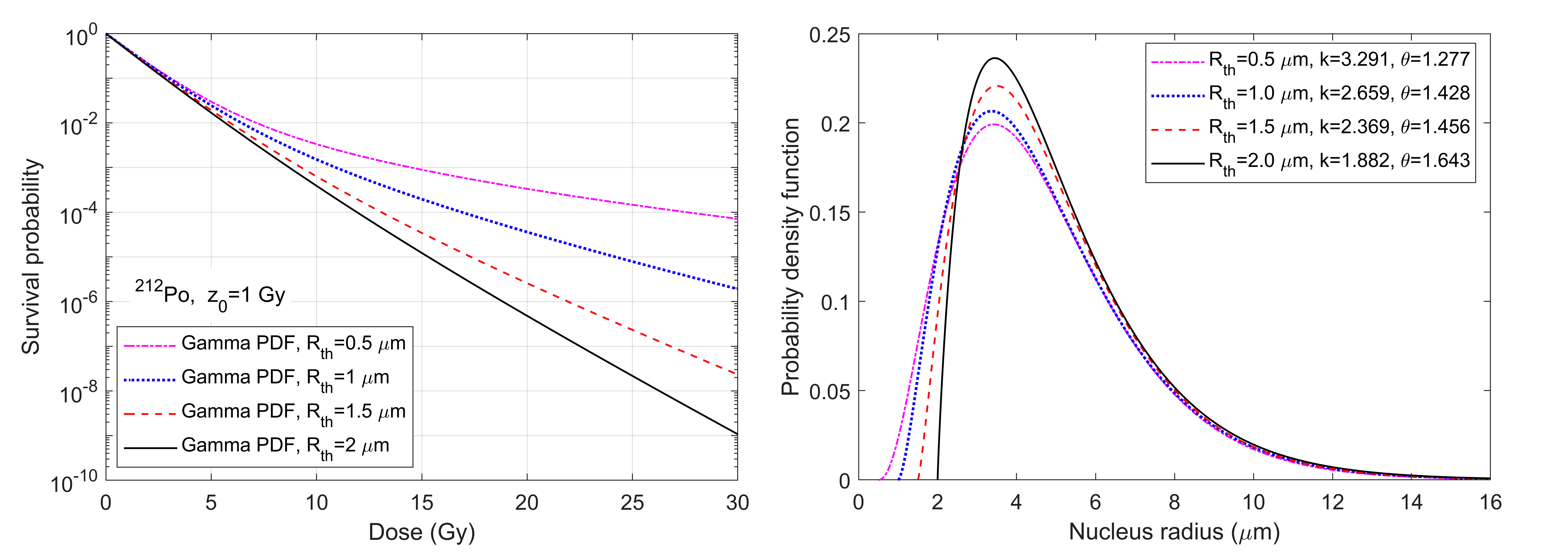}
   \caption{(Left) Survival probability plotted against dose for cells with nucleus radii sampled from the Gamma function-fitted distribution of Poole's data. The curves depict various thresholds $R_{th}$ for the minimum radius incorporated in the distribution, focusing on radioresistant cells ($z_0=1$~Gy) for $^{212}$Po alpha particles. (Right) Corresponding nucleus radius distributions.
   \label{fig:SP_Shirin} 
   }
    \end{center}
\end{figure}

Figure \ref{fig:SP_Shirin} displays survival curves, incorporating the impact of using Poole’s data while varying the threshold, $R_{th}$, over the range $0.5-2~\mu$m (with refitting and proper normalization of the Gamma PDF for each value of $R_{th}$). The calculation was performed for $^{212}$Po alpha particles and $z_0=1$~Gy. The deviation from exponential survival at high doses persists and is qualitatively the same as for the Gaussian PDF. Increasing $R_{th}$ has a similar result as reducing the width of the Gaussian PDF, with the survival curve progressively reverting to an exponential nature.

\begin{figure}[ht!]
   \begin{center}
   \includegraphics[width=0.6\textwidth]{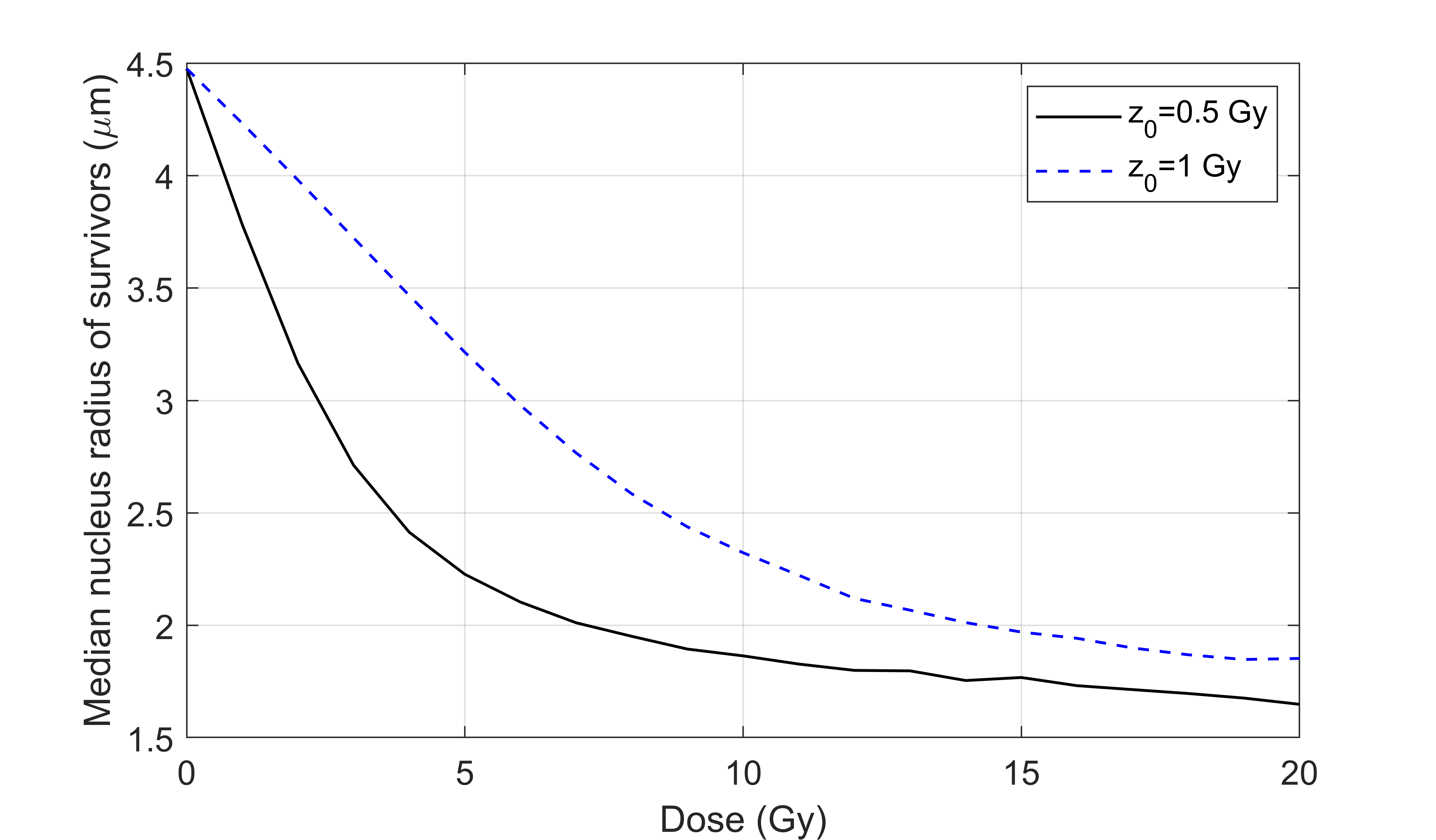}
   \caption{Median nucleus radius of surviving cells as a function of the absorbed alpha dose. The starting point is Poole's nucleus size distribution with a threshold nucleus radius of 1.5~$\mu$m. Cells are exposed to $^{212}$Po alpha particles. The two curves correspond to radiosensitive ($z_0=0.5$~Gy) and radioresistant ($z_0=1$~Gy) cells.}
   \label{fig:Survivors_Rn} 
    \end{center}
\end{figure}

\begin{figure}[ht!]
    \begin{center}
       \includegraphics[width=\textwidth]{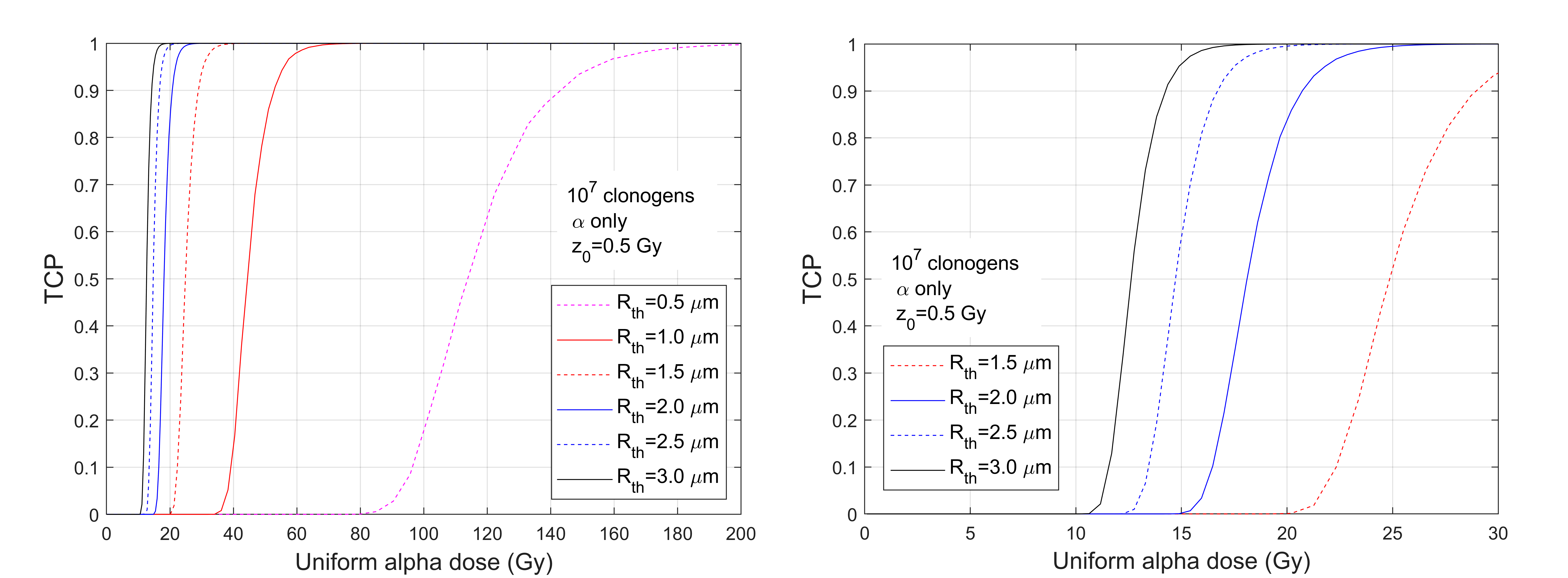}
       \caption{Effect of the nucleus size distribution threshold: TCP curves for a uniform dose field for varying threshold values on the minimal nucleus radius of Poole's data. The calculation includes only the alpha dose ($P_{leak}(Pb)=0.5$), and is performed for a tumor comprising $10^7$ clonogens, assuming $z_0$ = 0.5 Gy. The left panel shows the full dose range, while the right panel focuses on $0-30$~Gy.} 
       \label{fig:TCP_UNI_Rth} 
    \end{center}
\end{figure}

Figure \ref{fig:Survivors_Rn} examines further consequences of the large width of Poole's nucleus size distribution (more specifically, the inclusion of small nuclei). We consider here the case of a low threshold on the nucleus radius, $R_{th}=1.5~\mu$m, and show that as the dose increases, the median of the nucleus radius distribution decreases toward its threshold value. For high doses, surviving cells have similar nucleus radii (median of $\sim1.6-1.8~\mu$m), with a minor difference between radiosensitive ($z_0=0.5$~Gy) and radioresistant ($z_0=1$~Gy) cells. For each value of $z_0$ the curves are the average of 6 repetitions, each with $10^7$ cells. Fluctuations at high doses result from the very small number of surviving cells. Varying the threshold on the nucleus radius does not change the observation that as the dose increases, the median radius of the surviving cells decreases toward the threshold.
The impact of varying $R_{th}$ over the range $0.5-3~\mu$m on the TCP was calculated for a uniform dose distribution (with $P_{leak}(\mathrm{Pb})=0.5$) and $z_0=0.5$~Gy (Figure \ref{fig:TCP_UNI_Rth}, with the left panel showing the full range of dose values, and the right focusing on the region 0-30~Gy). The tumor volume is 10~cm$^3$ ($10^7$ clonogens, assuming $10^6$ clonogens/cm$^3$). Changing $R_{th}$ has a dramatic effect on the TCP curves. The dose corresponding to $TCP=50\%$, $D_{50}$, increases considerably from 12.6~Gy for $R_{th}=3~\mu$m to 18.1~Gy for $R_{th}=2~\mu$m. It then quickly diverges, reaching 44.5~Gy for $R_{th}=1~\mu$m and 113.8~Gy for $R_{th}=0.5~\mu$m. 

This is accompanied by a dramatic increase in the width of the rising section of the curve, resulting in an increase of $D_{90}$ (The dose corresponding to $TCP=90\%$) from 14.2~Gy for $R_{th}=3~\mu$m to 143~Gy for $R_{th}=0.5~\mu$m. Note that the strong effect stems entirely from the inclusion of small nuclei close to the threshold, and not from a change in the mean value of the nucleus radius; the latter presents a minor decrease from 5.6~$\mu$m for $R_{th}=3~\mu$m to 4.7~$\mu$m for $R_{th}=0.5~\mu$m. The extraordinary sensitivity of TCP to changes in $R_{th}$ for these small values emphasizes the critical importance of setting a realistic lower bound threshold in the nucleus size distribution of clonogenic cells. 

%===============================================================
\section{Discussion}
\vspace{0.5cm}
This study introduces a microdosimetric framework to investigate the impact of nucleus size variability on cell survival and tumor control probabilities in the context of Alpha-DaRT treatments. The findings of this study can provide new insights for Alpha-DaRT treatment planning, but are also relevant for other alpha-based treatment modalities.

The explicit incorporation of finite-width nucleus size distributions into the microdosimetric survival model reveals significant deviations from the exponential trend observed for uniform cell populations. This departure is most pronounced at high doses, where survival probabilities are dominated by cells with small nuclei (as larger cells tend to succumb first). This observation persists for both artificial Gaussian and empirically derived nucleus size distributions \parencite{Poole2015}, indicating its general nature. Furthermore, our analysis shows that the width of the nucleus size distribution strongly influences the survival gap between radiosensitive ($z_0$ = 0.5 Gy) and radioresistant ($z_0$ = 1 Gy) cells. Broader distributions include more small-nucleus cells, and as the width increases, the gap between the survival curves narrows, reflecting a reduced relative impact of $z_0$ on survival.  

Tumor control probability calculations reveal an extraordinary sensitivity to the lower threshold of the nucleus size distribution. Reducing the minimal nucleus threshold from $R_{th}$ = 3 $\mu$m to 0.5 $\mu$m dramatically increases $D_{90}$ (the dose required to achieve 90\% tumor control probability) from 14.2~Gy to 143~Gy. This sensitivity is consistent with previous observations reported in \parencite{SR2000}. This emphasizes the importance of accurate characterization of clonogenic nucleus sizes for reliable TCP modeling, particularly with respect to the smallest clonogen nucleus represented in the data. Otherwise, TCP may be severely underestimated. 

The inclusion of very small nuclei in the dataset may arise from genuine biological phenomena as well as methodological artifacts related to data analysis. One important biological source is the presence of \emph{micronuclei}, which are frequently observed in cancer cells \parencite{Crasta2012, DiBona2024}. Micronuclei are small, extranuclear bodies formed in the cytoplasm as a consequence of chromosome mis-segregation during cell division. They consist of chromosome fragments or entire chromosomes that were not incorporated into the primary nucleus. Given the documented fates of micronuclei, including persistence, degradation, or reincorporation into the primary nucleus of a daughter cell \parencite{Crasta2012, Reimann2023}, micronuclei do not appear to function as independent reproductive units. A defining feature of micronuclei is their small size—--typically 1–5 $\mu$m in diameter---depending on the cell type \parencite{Kneissig2019, Reimann2023}. Because they contain condensed chromatin, micronuclei stain similarly to intact nuclei in hematoxylin and eosin staining (H\&E), making them indistinguishable in standard histological analyses. Consequently, the size distribution used in this study may include micronuclei, which do not possess reproductive potential. To avoid their inclusion in future studies, image processing algorithms can be designed to identify the main nucleus of each cell and exclude neighboring micronuclei (including only the main nucleus in the distribution), or discard cells with micronuclei altogether.

Another biological factor that may lead to the presence of very small nuclei in the data is mechanical deformation during cell migration through dense extracellular environments. Since the nucleus is typically the largest and stiffest organelle, it often becomes the main physical barrier during migration through narrow spaces. When the constrictions are smaller than the resting nuclear diameter, the nucleus undergoes pronounced compression and elongation to allow translocation---a well-documented phenomenon in cancer cell migration through confined environments \parencite{Wolf2013, Golloshi2022}. Experimental studies have shown that nuclei can transiently adopt minimal dimensions of approximately 2–3 $\mu$m along one axis under severe confinement \parencite{Denais2016, Shah2021}. 

A methodological source of uncertainty arises from the use of histological data that do not distinguish between clonogenic and non-clonogenic cancer cells. The nucleus size data used in this study were derived from a method that reconstructs three-dimensional cell distributions from two-dimensional histological sections \parencite{Poole2015}. Because the present microdosimetric model quantifies TCP in terms of clonogenic cells, it is particularly sensitive to their nucleus size distribution. However, H\&E staining, if used as a stand-alone technique, does not differentiate clonogenic from non-clonogenic populations, suggesting that both may be represented in the dataset. Based on literature searches, the possibility that non-clonogenic cells exhibit smaller nuclei cannot be excluded, as the relationship between nuclear size and clonogenic potential is influenced by tumor type, cell-cycle stage, and microenvironmental factors.

Lastly, the authors of \cite{Poole2015} mention that the morphological reconstruction technique may be subject to inherent methodological limitations; deviations between histological and true in-vivo cell arrangements were not considered, and the fixation and staining procedures used in tissue preparation can cause cell distortion or shrinkage. In addition, the statistical uncertainties associated with the size extraction algorithm, as well as the packing fraction of the reconstructed three-dimensional models, have not yet been quantified. Such effects could contribute to the appearance of artificially small nuclei within the reconstructed dataset, leading to an extended lower tail of the size distribution. 

Several studies have reported nucleus size measurements for different cancer cell lines. However, these values generally reflect the entire cell population rather than clonogenic subpopulations, underscoring the need for techniques that can distinguish clonogenic from non-dividing cancer cells. Reported ranges of nucleus radii include $\sim$3.2–4.7 $\mu$m for breast cancer \parencite{Kashyap2017} (derived from equivalent diameter estimations under the assumption of perfectly spherical nuclei), and 3.1–5.1 $\mu$m for prostate cancer \parencite{Malshy2022} (based on minimal and maximal nucleus radius measurements). It is important to note that some of the techniques used in these studies may cause nucleus shape deformation, as cells placed on slides tend to flatten into 2D disc-like shapes. As a result, the measured nucleus radius in the flattened plane may appear larger than it would in the native 3D configuration. An extensive literature did not result in lower empirical values for solid tumor nuclei. Given the scarcity of direct evidence on clonogen-specific nucleus sizes, a conservative lower threshold between 2 and 3~$\mu$m can be adopted to constrain uncertainty in TCP estimates, with the understanding that this range may also include unrealistically small and potentially non-clonogenic nuclei. The difference in $D_{90}$ between $R_{th}$ = 2~$\mu$m and 3~$\mu$m is approximately 7~Gy---still significant but substantially smaller than the uncertainty introduced by including nuclei with radii smaller than 2~$\mu$m. 

While the present study considered uniform dose fields, a separate publication will incorporate realistic intra-tumoral dose gradients generated by Alpha-DaRT source arrangements. Such extensions will allow for a more comprehensive assessment of TCP under clinically relevant dose distributions. Additionally, exploring the interplay between nucleus size distributions and other sources of heterogeneity, such as clonogen density and micro-environmental factors, may further refine the predictive tools.

The integration of realistic nucleus size variability into microdosimetric calculations represents an important step toward improving the accuracy of TCP predictions for Alpha DaRT and other alpha-based therapies. Continued methodological and biological refinements will support more robust treatment planning and strengthen the clinical use of microdosimetric models.

%===============================================================
\section{Conclusions}
\vspace{0.5cm}

This work highlights the critical role of nucleus size variability in shaping the biological response of tumors to alpha-particle irradiation. By explicitly incorporating finite-width nucleus size distributions into survival and TCP models, we demonstrate significant deviations from classical exponential survival and reveal the strong influence of small-nucleus subpopulations on predicted tumor control probabilities. These effects are not directly evident when assuming uniform nucleus size distributions. In particular, the inclusion of unrealistically small nuclei leads to a significant overestimation of the required dose for tumor control. Future work should aim to refine nucleus size characterization and improve the biological relevance of the data. This includes identifying clonogenic versus non-clonogenic nuclei within the cell population, differentiating micronuclei from main nuclei during data extraction to reduce background contributions, and measuring clonogen-specific size distributions. Establishing a reliable threshold or threshold range based on these measurements and incorporating it into the microdosimetric TCP model would significantly improve predictive accuracy.

\funding{This work was partly supported by Alpha Tau Medical Ltd.}

\data{Data included in this article are available upon request from L. Arazi.}

\vspace{0.5cm}
\noindent\textbf{ORCID iDs}
\newline
\noindent Yevgeniya Korotinsky 0000-0003-4075-8919
\vspace{0.05cm} \newline
\noindent Lior Arazi 0000-0002-7624-5827

\printbibliography
\end{document}